\documentclass[12pt]{article}
\usepackage[latin1]{inputenc}
\usepackage[dvips]{graphicx}
\usepackage{graphicx}
\newcommand{\exprm}{{\rm exp}}

\newcommand{\text}{\rm}

\newcommand{\ug}{ \; = \; }

\newcommand{\bb}{\begin{equation}}
\newcommand{\ee}{\end{equation}}
\newcommand{\bega}{\begin{eqnarray}}
\newcommand{\ega}{\end{eqnarray}}
\newcommand{\begae}{\begin{eqnarray*}}
\newcommand{\egae}{\end{eqnarray*}}

\newcommand{\h}{\hspace*{4ex}}
\newcommand{\dis}{\displaystyle}

\newcommand{\be}{\beta}

\newcommand{\om}{\omega}

\newcommand{\cent}{\centerline}
\newcommand{\vs}{\vspace*}

\begin{document}

\baselineskip 1.0cm

\begin{center}

{\Large {\bf Diffraction-Attenuation Resistant Beams \\ in
Absorbing Media}$^{\: (\dag)}$} \footnotetext{$^{\: (\dag)}$  Work
partially supported by FAPESP (Brazil).\\ E-mail address for
contacts: mzamboni@dmo.fee.unicamp.br}

\end{center}

\baselineskip 0.8cm

\vs{5mm}

\cent{ Michel Zamboni-Rached, }

\vs{0.2 cm}

\centerline{{\em DMO--FEEC, State University at Campinas,
Campinas, SP, Brazil.}}

\begin{abstract}
In this work, in terms of suitable superpositions of
equal-frequency Bessel beams, we develop a theoretical method
to obtain nondiffractive beams in (weakly conductive)
{\it absorbing media} capable to resist the loss effects for long
distances.
\end{abstract}


{\em PACS nos.}: \ 41.20.Jb ; \ 03.50.De ; \ 03.30.+p ; \ 84.40.Az
; \ 42.82.Et ; \ 83.50.Vr ; \ \ 62.30.+d ; \ 43.60.+d ; \
91.30.Fn ; \  04.30.Nk ; \  42.25.Bs ; \ 46.40.Cd ; \ 52.35.Lv \
.\hfill\break

\section{Introduction}

\h Over many years, the theory of nondiffracting beams and pulses
has been developed, generalized and experimentally verified in
many fields, such as optics, microwaves and acoustics[1-11].

\h When propagating in a non-absorbing medium, these waves
maintain their spatial shape for long distances, i.e., they
possess a large depth of field.

\h However, the situation is not the same when dealing with
absorbing media. In these cases, both the ordinary and the
nondiffracting beams (and pulses) will suffer the same effect: an
exponential attenuation along the propagation axis.

\h Here, we are going to show that, through suitable
superpositions of equal-frequency Bessel beams, it is possible to
obtain nondiffracting beams {\it in absorbing media} capable to
resist the loss effects, maintaining amplitude and spot
size of their central core for long distances.

\h The method that will be developed here is a generalization of a
previous one (also developed by us)\cite{Mic,Mic2}, which was
conceived for lossless media, where suitable superpositions of
Bessel beams are used to construct {\it stationary} wave fields
with arbitrary longitudinal intensity shape. Those new solutions
were called Frozen Waves.

\h Before continuing, some points are to be stressed: (a) the
method developed here addresses to poorly conductive media
($\sigma/\varepsilon\om <<1$) in the optical frequency range;
actually, we are considering materials where the penetration depth
of the optical beams are typically $> 10^{-4}$m; \  (b) the energy
absorption by the medium continues to occur normally, the
difference being in that these new beams have an initial
transverse field distribution, such to be able to reconstruct (even in
the presence of absorption) their central cores for distances
considerably longer than the penetration depths of ordinary
(nondiffracting or diffracting) beams.

\h In the next Section we shall extend the previous
method\cite{Mic,Mic2} to absorbing media. Section 3 is devoted to
obtaining examples of these new beams in some interesting
situations.

\section{The Mathematical Methodology}

\h The method that will be developed in this Section is based on
Bessel beams superpositions; so, it is appropriate first to
recall some characteristics of such beams in absorbing media.

\h In the same way as for lossless media, we construct a Bessel
beam in the absorbing materials by superposing plane waves whose wave
vectors lie on the surface of a cone with vertex angle $\theta$.
The refractive index of the medium can be written as $n(\om) =
n_R(\om) + in_I(\om)$, quantity $n_R$ being the real part of the complex
refraction index and $n_I$ the imaginary one, responsible for the
absorbing effects. With a plane wave, the penetration depth
$\delta$ for the frequency $\om_0$ is given by $\delta =
c/2\om_0 n_I$.

\h In this way, a zero-order Bessel beam in dissipative media can
be written as \\
 $\psi = J_0(k_{\rho}\rho){\rm exp}(i\beta z){\rm
exp}(-i\om t)$ with $\beta = n(\om)\om \cos\theta/c = n_R\om
\cos\theta/c + in_I\om \cos\theta/c \equiv \beta_R + i\beta_I$;
$k_{\rho}=n_R\om \sin\theta/c + in_I\om \sin\theta/c \equiv
k_{\rho R} + ik_{\rho I}$, and so $k_{\rho}^2 = n^2\om^2/c^2 -
\beta^2$. In this way $\psi=J_0((k_{\rho R} + ik_{\rho
I})\rho){\rm exp}(i\beta_R z){\rm exp}(-i\om t){\rm exp}(-\beta_I
z)$, where $\beta_R$, $k_{\rho R}$ are the real parts of the
longitudinal and transverse wave numbers, and $\beta_I$, $k_{\rho
I}$ are the imaginary ones, while the absorption coefficient of a
Bessel beam with an axicon angle $\theta$ is given by
$\alpha_{\theta}=2\beta_I=2n_I\om \cos\theta/c$, its penetration
depth being $\delta=1/\alpha=c/2\om n_I\cos\theta$.

\h Due to the fact that $k_{\rho}$ is complex, the amplitude of the
Bessel function
$J_0(k_{\rho}\rho)$ starts decreasing from $\rho=0$
till the transverse distance $\rho=1/2k_{\rho_I}$, and afterwards it
starts growing exponentially.  This behavior is not
physically acceptable, but one must remember that it occurs only because
of the fact that an ideal Bessel beam needs an infinite aperture to
be generated.
However, in any real situation, when a Bessel beam is generated by
finite apertures, that exponential growth in the transverse
direction, starting after $\rho=1/2k_{\rho_I}$, will {\it not} occur
indefinitely, stopping at a given value of $\rho$. Actually, taking
into account the size of the aperture generally used for generating
optical Bessel beams, and the fact that we are considering media
that are poor conductors, that exponential growth does not even happen,
and the resulting Bessel beam only presents a decreasing intensity
in the transverse direction.

\h Let us now present our method.

\h Consider an absorbing medium with the complex refraction index
$n(\om) = n_R(\om) + in_I(\om)$, and the following superposition
of $2N + 1$ Bessel beams with the same frequency $\om_0$

\bb \dis{\Psi(\rho,z,t) \ug \sum_{m=-N}^{N} A_m\,J_0\left((k_{\rho
R_m} + ik_{\rho
I_m})\rho\right)\,e^{i\,\be_{R_m}z}\,e^{-i\,\om_0\,t}\,e^{-\be_{I_m}z}
} \; , \label{soma1} \ee

\

where $m$ are integer numbers, $A_m$ are constant (yet unknown)
coefficients, $\be_{R_m}$ and $k_{\rho R_m}$ ($\be_{I_m}$ and
$k_{\rho I_m}$) are the real parts (the imaginary parts) of the
complex longitudinal and transverse wave numbers of the m-th
Bessel beam in superposition (\ref{soma1}), the
following relations being satisfied

\bb
 k_{\rho_m}^2 \ug n^2\frac{\om^2}{c^2} - \be_{m}^2 \label{kr}
\ee

\bb \frac{\be_{R_m}}{\be_{I_m}} \ug \frac{n_R}{n_I} \label{bei}
\ee

\bb \frac{k_{\rho R_m}}{k_{\rho I_m}} \ug \frac{n_R}{n_I}
\label{ki} \ee

\

where $\be_m = \be_{R_m} + i\be_{I_m}$ and $k_{\rho_m} = k_{\rho
R_m} + ik_{\rho I_m}$.

\h Our goal is now to find out the values of the longitudinal
wave numbers $\be_m$ and the coefficients $A_m$ in order to
reproduce approximately, inside the interval $0 \leq z \leq L$ (on
the axis $\rho=0$), a {\it freely chosen} longitudinal intensity pattern that
we call $|F(z)|^2$.

\h The problem was already solved by us and the method developed for the
particular case of lossless media\cite{Mic,Mic2}, i.e., when
$n_I=0 \rightarrow \be_{I_m}=0$. For those cases, it was shown
that the choice $\beta=Q+2\pi m/L$, with $A_m = \int_{0}^{L}
F(z){\rm exp}(-i2\pi mz/L)/L\,\,dz$ can be used to provide
approximately the desired longitudinal intensity pattern
$|F(z)|^2$ on the propagation axis, within the interval $0\leq z
\leq L$, and, at same time, to regulate the spot size of the
resulting beam by means of the parameter $Q$, which can be also used to
obtain large field depths (see details in Refs.\cite{Mic,Mic2}).

\h However, when dealing with absorbing media, the procedure
described in the last paragraph does not work, due to the presence of
the functions $\exprm (-\be_{I_m}z)$ in the superposition
(\ref{soma1}), since in this case that series does not became a
Fourier series when $\rho=0$.

\h On attempting to overcome this limitation, let us write the real
part of the longitudinal wave number, in superposition (\ref{soma1}), as

\bb \be_{R_m} \ug Q + \frac{2\pi m}{L} \label{br} \ee

with

\bb 0 \leq Q + \frac{2\pi m}{L} \leq n_R \frac{\om_0}{c}
\label{cond} \ee

\

where this inequality guarantees forward propagation only, with no
evanescent waves.

\h In this way the superposition (\ref{soma1}) can be written

\bb \dis{\Psi(\rho,z,t) \ug e^{-i\,\om_0\,t}\,e^{i\,Qz}\,
\sum_{m=-N}^{N} A_m\,J_0\left((k_{\rho R_m} + ik_{\rho
I_m})\rho\right)\,e^{i\,\frac{2\pi m}{L}z}\,e^{-\be_{I_m}z} } \; ,
\label{soma2} \ee

\

where, by using Eqs.(\ref{bei}), quantity $\be_{I_m}$ is given by

\bb \be_{I_m} \ug \frac{n_I}{n_R}\be_{R_m} \ug
\frac{n_I}{n_R}\left(Q + \frac{2\pi m}{L} \right) \label{bi2} \ee

\

and $k_{\rho_m}=k_{\rho R_m} + ik_{\rho I_m}$ is given by
Eq.(\ref{kr}).

\h Now, let us examine the imaginary part
of the longitudinal wave numbers. From Eq.(\ref{bi2}) the minimum
and maximum values among the $\be_{I_m}$ are $(\be_I)_{\rm
min}=(Q-2\pi N/L)n_I/n_R$ and $(\be_I)_{\rm max}=(Q+2\pi
N/L)n_I/n_R$, the central one being given by $\overline{\be}_I
\equiv (\be_I)_{m=0} = Q n_I/n_R $. With this in mind, let us evaluate the
ratio

\bb \Delta = \frac{(\be_I)_{\rm max} - (\be_I)_{\rm
min}}{\overline{\be}_I} = \frac{4 \pi N}{L Q} \label{cond2} \; . \ee

\

\h Thus, when $\Delta <<1$, there are no considerable differences
among the various $\be_{I_m}$, since it holds $\be_{I_m} \approx
\overline{\be}_I$ for all $m$. And, in the same way, there are no
considerable differences among the exponential attenuation
factors, since $\exprm (-\be_{I_m}z) \approx \exprm
(-\overline{\be}_I z)$. So, when $\rho=0$ the series in the r.h.s.
of Eq.(\ref{soma2}) can be approximately considered a truncated
Fourier series {\it multiplied by} the function $\exprm
(-\overline{\be}_I z)$:  and, therefore, superposition (\ref{soma2})
can be used to reproduce approximately the desired longitudinal
intensity pattern $|F(z)|^2$ (on $\rho=0$), within $0\leq z \leq
L$, when the coefficients $A_m$ are given by

\bb A_m \ug \frac{1}{L}\,\int_{0}^{L} F(z)\,e^{\overline{\be}_I
z}e^{-i\,\frac{2\pi m}{L}z}\,dz \label{am} \ee

\

the presence of the factor $\exprm (\overline{\be}_I z)$ in
the integrand being necessary to compensate for the factors $\exprm
(-\be_{I_m}z)$ in superposition (\ref{soma2}).

\h Since we are adding together zero-order Bessel functions, we
can expect a good field concentration around $\rho=0$.

\h In short, we have shown in this Section how one can get, in a
weakly conductive {\it absorbing medium},  a {\it stationary} wave-field
with a good transverse concentration, and whose longitudinal
intensity pattern (on $\rho=0$) can approximately assume any
desired shape $|F(z)|^2$ within the predetermined interval $0 \leq
z \leq L$. The method is a generalization of a previous
one\cite{Mic,Mic2} and consists in the superposition of Bessel beams in
(\ref{soma2}), the real and imaginary parts of their longitudinal
wave numbers being given by Eqs.(\ref{br})and (\ref{bi2}), while
their complex transverse wave numbers are given by Eq.(\ref{kr}),
and, finally, the coefficients of the superposition are given by
Eq.(\ref{am}). The method is justified, since $4\pi N/LQ << 1$;
happily enough, this condition is satisfied in a great number of
situations.

\h In the next Section we shall apply this method for obtaining new
beams whose central core can resist the effects caused by
diffraction and attenuation in absorbing media for long distances.

\section{Examples: Almost Undistorted Beams in Absorbing Media}

\h Let us consider a biological tissue as the absorbing medium.  In
the wavelength range $280-300$nm, proteins are the major absorbers,
being the UV absorption similar for many types of tissues.

\h A typical XeCl excimer laser ($\lambda = 308 {\rm nm}
\rightarrow \om = 6.09\times 10^{15}$Hz) has a penetration depth,
in this case, of 5 cm; i.e., an absorption coefficient $\alpha =
20 {\rm m}^{-1}$, and therefore $n_I = 0.49\times 10^{-6} $. The
real part of the refraction index for this wavelength can be
approximated by $n_R = 1.5$, and therefore $n = n_R + in_I = 1.5 +
i\,0.49\times 10^{-6}$.

\h A typical Bessel beam with $\om_0  = 6.09\times 10^{15}$Hz and
with an axicon angle $\theta = 0.0141$rad (so, with a transverse
spot of radius $8.4\,\mu$m), when generated by an aperture, say,
of radius $R=3.5\;$mm, can propagate in vacuum a distance (its
field depth) equal to $Z=R/\tan\theta=25\;$cm while resisting the
diffraction effects. However, in the material medium considered
here, the penetration depth of this Bessel beam would be only $z_p
= 5\;$cm.

\h Now, we can use the method of the previous Section to obtain,
in the same medium and for the same wavelength, an almost
undistorted beam capable of preserving its spot size and the
intensity of its {\it central core} for a distance many times
larger than the typical penetration depth of an ordinary beam
(nondiffracting or not).

\h With this purpose, let us suppose that, for this material
medium, we want a beam (with $\om_0  = 6.09\times 10^{15}$Hz) that
maintains amplitude and spot size of its central core for a
distance of $25\;$cm, i.e., a distance 5 times greater than the
penetration depth of an ordinary beam with the same frequency. We can
model this beam by choosing the desired longitudinal intensity
pattern $|F(z)|^2$ (on $\rho=0$), within $0 \leq z \leq L$,

 \bb
 F(z) \ug \left\{\begin{array}{clr}
&1 \;\;\; {\rm for}\;\;\; 0 \leq z \leq Z  \\

 \\

&0 \;\;\; \mbox{elsewhere} ,
\end{array} \right.  \label{Fz}
 \ee

\

and by putting $Z=25\;$cm, with, for example, $L=33\;$cm.

\h Now, the Bessel beam superposition (\ref{soma2}) can be used to
reproduce approximately this intensity pattern, and to this
purpose let us choose $Q=0.9999\om_0 / c$ for the $\be_{R_m}$ in
Eq.(\ref{br}), and $N=20$ (which is allowed by inequality
(\ref{cond})).

\h Once we have chosen the values of $Q$, $L$ and $N$, the
values of the complex longitudinal and transverse Bessel beams
wave numbers happen to be defined by relations (\ref{br}), (\ref{bi2}) and
(\ref{kr}). Eventually, we can use Eq.(\ref{am}) and find out the
coefficients $A_m$ of the fundamental superposition (\ref{soma2}),
that defines the resulting stationary wave-field.

\h Let us just note that the condition $4\pi N/LQ << 1$ is
perfectly satisfied in this case.

\h In Fig.1(a) we compare the desired longitudinal intensity
function $|F(z)|^2$ with the resulting stationary wave field,
obtained from the Bessel beam superposition (\ref{soma2}), and we
can notice a good agreement between them. A better result could be
reached by using a higher value of $N$, which in this specific
example may assume, according to inequality (\ref{cond}), a
maximum value of $158$.

\h It is interesting to note that. at this distance (25 cm), an
ordinary beam would have got its initial field-intensity attenuated
$148$ times.

\

\begin{figure}[!h]
\begin{center}
 \scalebox{1}{\includegraphics{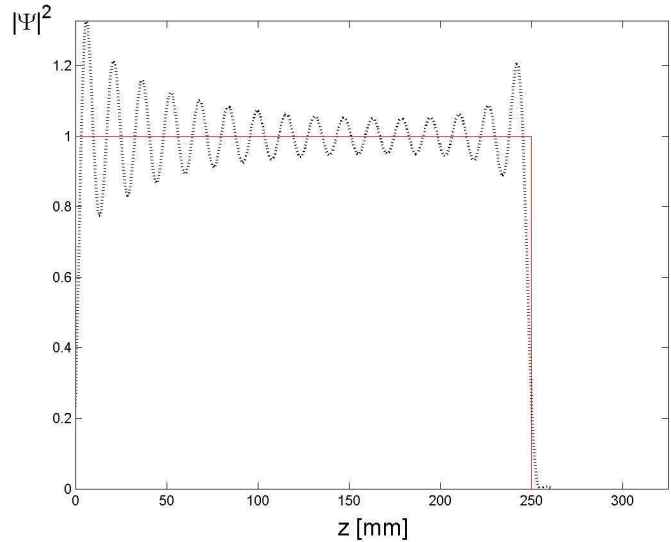}}
\end{center}
\caption{Comparison between the intensity of the desired
longitudinal function $F(z)$ (solid line) and that obtained from
the Bessel beam superposition (\ref{soma2}) (dotted line). One can
observe a good agreement between them.} \label{fig1}
\end{figure}

\

\h Figure 2(a) shows the 3D field-intensity of the resulting beam.
One can see that the field possesses a good transverse
localization (with a spot size smaller than $10 \mu$m), it being
capable of maintaining spot size and {\it intensity} of its
central core till the desired distance. Figure 2(b) shows the same
picture, but in an orthogonal projection.


\

\begin{figure}[!h]
\begin{center}
 \scalebox{.9}{\includegraphics{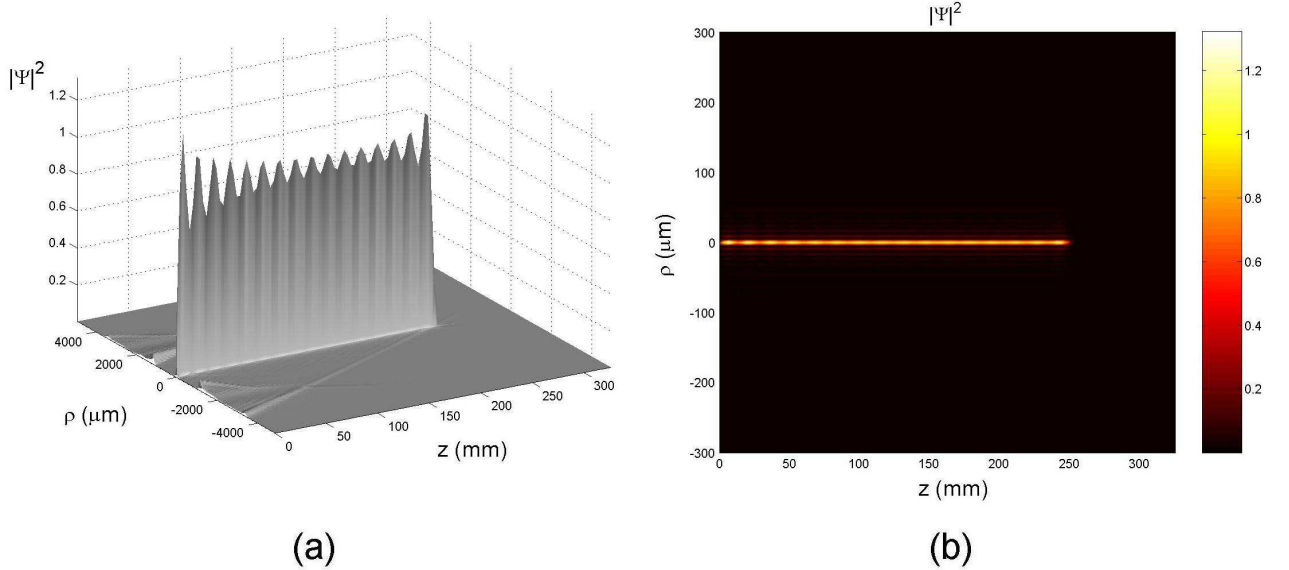}}
\end{center}
\caption{\textbf{(a)} Three-dimensional field-intensity of the
resulting beam. \textbf{(b)} The same figure in orthogonal
projection.} \label{fig2}
\end{figure}

\

\h As we have said in the Introduction, the energy absorption by
the medium continues to occur normally; the difference is that
these new beams have an initial transverse field distribution
sophisticated enough to be able to reconstruct (even in the
presence of absorption) their central cores till a certain
distance. For a better visualization of this field-intensity
distribution and of the energy flux, Fig.3(a) shows the resulting
beam, in an orthogonal projection and in logarithmic scale. It is
clear that the energy comes from the lateral regions, in order to
reconstruct the central core of the beam. Figure 3(b) presents the
initial field-intensity distribution on the aperture plane, at
$z=0$. We can note the complicated initial transverse
field-intensity distribution needed for feeding the central core
of the beam along the propagation axis.

\begin{figure}[!h]
\begin{center}
 \scalebox{.9}{\includegraphics{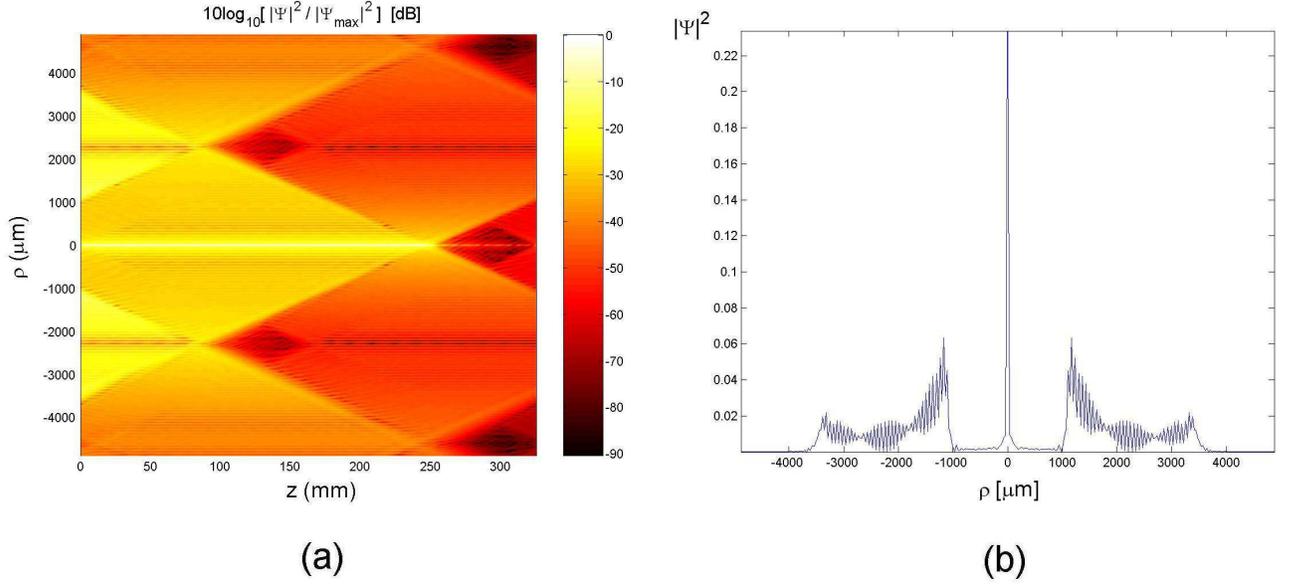}}
\end{center}
\caption{\textbf{(a)} The resulting beam, in an orthogonal
projection and in logaritmic scale. \textbf{(b)} Initial
field-intensity distribution on the aperture plane, at $z=0$.} \label{fig3}
\end{figure}

\h Obviously, the amount of energy necessary to construct these
new beams is greater than that necessary to generate an ordinary
beam in a non-absorbing medium.

\h And it is also clear that there is a {\it limitation} on the depth of
field of these new beams. For distances longer than 8 or 10 times
the penetration depth of an ordinary beam, besides a great energy
demand, we meet the fact that the field-intensity in the lateral regions
would be even
higher than that of the core, and the field would loose
the usual characteristics of a beam (transverse field
concentration).

\section{Conclusion}

\h In this paper we have shown that it is possible to construct
almost-undistorted beams in absorbing media. They are got by
suitable Bessel beams superpositions, and are capable to resist the
loss effects, maintaining amplitude and spot size of their
central core for long distances, when compared with the usual
penetration depths of ordinary beams.

\section{Acknowledgements}

The author is very grateful to Erasmo Recami, Hugo E.
Hern\'andez-Figueroa and Claudio Conti for continuous discussions
and kind collaboration.

\end{document}